# Assessing AI Explainability: A Usability Study Using a Novel Framework Involving Clinicians

Mohammad Golam Kibria
*Carolina Health Informatics Program*
*University of North Carolina at Chapel Hill*
North Carolina, USA
kibria@email.unc.edu

Lauren Kucirka
*UNC Maternal Fetal Medicine*
*University of North Carolina at Chapel Hill*
North Carolina, USA
lauren_kucirka@med.unc.edu

Javed Mostafa
*Faculty of Information*
*University of Toronto*
Toronto, Canada
dr.javedm@utoronto.ca

*Abstract*—**An AI design framework was developed based on three core principles, namely understandability, trust, and usability. The framework was conceptualized by synthesizing evidence from the literature and by consulting with experts. The initial version of the AI Explainability Framework was then validated based on an in-depth expert engagement and review process. For evaluation purposes, an AI-anchored prototype, incorporating novel explainability features, was built and deployed online via Google Cloud Platform. The primary function of the prototype was to predict the postpartum depression risk using analytics models. The development of the prototype was carried out in an iterative fashion, based on a pilot-level formative evaluation, followed by a round of refinement and summative evaluation. In the formative stage, the prototype was evaluated based on an internal pilot usability test involving a small number of clinicians (n=3). The prototype was updated based on the user's feedback in the formative stage. The System Explainability Scale (SES) metric was developed to measure the individual and interacting influence of the three dimensions of the AI Explainability Framework. For the summative stage, a comprehensive usability test was conducted involving 20 clinicians, and the SES metric was used to assess clinicians' satisfaction with the tool. On a 5-point rating system, the tool received high scores for usability dimension, followed by trust and understandability. The average explainability score was 4.56. In terms of understandability, trust and usability, the average score was 4.51, 4.53, and 4.71 respectively. Overall, the 13-item SES metric showed strong internal consistency with Cronbach's alpha of 0.84 and a positive correlation coefficient (Spearman's rho = 0.81, p<0.001) between the composite SES score and explainability, indicating a positive trend in AI explainability. This study demonstrated the influence of understandability, trust, and usability on AI Explainability using a combination of a novel design and experimental approach. A major finding was that the AI Explainability Framework, combined with the SES usability metric, provides a straightforward yet effective approach for developing AI-based healthcare tools that lower the challenges associated with explainability.**

## I. Introduction

In recent years, artificial intelligence (AI) is revolutionizing and transforming every sector, including healthcare, as part of the fourth industrial revolution or health 4.0 [1]. Numerous changes in several sub-domains of healthcare have been noticed, including maternal health [2]. The health sector is producing an enormous amount of data due to the widespread adoption of electronic medical records (EMR) systems. The EMR database contains health data in a variety of formats. These formats include structured, semi-structured, or unstructured clinical, genetic, and imaging data. Clinicians often struggle to effectively utilize such a massive amount of data for clinical decision making. AI technologies such as machine learning (ML) have the potential to revolutionize the decision-making processes in healthcare. For example, AI tools can handle large volumes of data, detect subtle patterns that humans may overlook, and produce highly accurate prediction results. These results are derived by utilizing machine-learning-based white-box and black-box models, including super learners [3]. In many situations, black-box models outperform traditional statistical or white-box linear models such as logistic regression, which assumes each risk factor impacts the outcome at the same level. However, in certain clinical contexts, logistic regression may be advantageous due to its interpretability, particularly in scenarios involving smaller datasets or when the model's influence on clinical decision making requires comprehensive transparency [4]. White-box models have simple rules and limited parameters, enabling humans to understand their underlying calculations. Conversely, black-box models use hundreds or even thousands of decision trees, e.g. "random forests", or billions of parameters in deep learning models, making it challenging for humans to understand their decisions [5]. A small number of studies have demonstrated the utility of these models, thereby highlighting their potential in improving clinical practice [6]. However, the major gap remains in the explainability of such models in clinical risk assessment [7]. Utilizing black-box algorithms in the medical field has also raised concerns among clinicians due to their opaqueness and lack of trustworthiness [8]. In many scenarios, clinicians will not trust an algorithm that lacks external validation and has output that is not easily explainable [9] or not medically contextualized [10]. It is essential to have healthcare providers' trust in the data and understand the reasoning behind predictions, since their decisions have potential clinical consequences [11]. The above issues demonstrate that AI-explainability remains at the core of the challenges despite extensive research in this field [12].

Since the USA's Defense Advanced Research Projects Agency (DARPA) was launched, there has been a notable increase in interest in the problem of ML explanation [13]. The demand for algorithmic accountability resulted in a regulatory framework, as evidenced by the European Union's

General Data Protection Regulation (GDPR). This declares the "right to explanation" and mandates AI systems to provide rationales for algorithmic decisions based on the user's request [14]. Meanwhile, literature searches suggest that several articles have been published on explainable AI in the past few years in different domains, including healthcare [11], [15], [16]. Following the GDPR directives, the EU recently enacted the AI Act to address the growing legal and ethical imperative for transparency in AI, especially in high-risk sectors like healthcare [17].

The rapid expansion and innovation in this emerging field saw a tremendous effort from researchers on method development [18], [19], [20], [21], [22], [23]. However, the community has recently started to draw attention to some of the shortcomings, including the absence of clear goals on explainability, inconsistent language, and disagreement over methods and metrics for assessing the quality of explanations [24]. At the same time, a few studies also presented conceptual or theoretical frameworks. Such frameworks were developed based on the user's contextual awareness [25], theory-driven or human reasoning aspects [26], or mental models [27], etc. However, not all of these studies applied any empirical evaluation of these concepts.

While previous studies have examined discrete constructs like ethics, bias, and fairness [28], trustworthiness [9], transparency [16], interpretability [16], [29] explainability [30], [21], [31] meaningfulness [32] or understandability [16], [31]. There is a lack of consensus on how these wide-ranging constructs should be systematically organized, orthogonalized, defined, and measured within a unified framework. The majority of the published explainable AI (XAI) studies are conceptual in nature and lack empirical evidence that demonstrates users actively engaged in evaluating XAI-based risk assessment tools following a specific AI Explainability Framework [33]. There are many constructs with considerable variability in their definitions, conceptual or theoretical frameworks, and metrics within explainability research, which may be challenging to comprehend by busy clinicians. The main intent of this study is to propose a streamlined framework and metrics as a potential solution to make explainability research in healthcare more meaningful and actionable in real-world contexts.

Drawing from the challenges and gaps identified above, the following research questions were formulated:

1) What are the key dimensions that anchor the explainability of AI tools in the contexts of healthcare?
2) Can the effectiveness of these dimensions be measured through a system-based usability study involving clinicians?

## II. METHODOLOGY AND EVALUATION PLAN

The above questions served as guides for a deductive methodology [34] utilizing the pre-existing knowledge, theories, and conceptual models on explainable AI as documented in scholarly literature. A user-centered design methodological approach, namely the "Double Diamond Design Framework", was formulated to examine the two research questions [35]. To achieve this goal, a concept analysis based on the user's "chain of logic" within their mental model was performed as a step towards explainability. This approach facilitated systematically identifying and defining the key dimensions of the conceptual framework. The construction of the framework was informed by a rigorous literature analysis and a multidisciplinary expert discussion panel that consisted of information scientists, medical doctors, and an evaluation specialist. The key dimensions of the framework were selected based on their relevance to directly addressing the explainability challenges, frequency of discussion in scholarly articles, and its applicability to real-world healthcare scenarios. The proposed dimensions – understandability, trust, and usability – were prioritized due to their distinct yet interconnected aspects of explainability, addressing both the technical and user-centric challenges recognized in the literature. The methodology anchored itself in a conceptual framework, namely "AI Explainability Framework" (Fig. 1), that identified the dimension of explainability. This framework was utilized to develop an AI prototype system and its usability evaluation methods. Furthermore, a metric was introduced, the System Explainability Scale (SES), to measure the effectiveness of each dimension of the explainability framework in relation to an AI-augmented clinical decision support system for real-time patient risk prediction.

The main objective of the study was to propose a straightforward framework and validate it by clinicians, thereby assessing the framework's applicability or effectiveness in healthcare. This usability study enhanced a first-hand and deeper understanding of clinicians' expectations from AI-driven tools and factors that may foster confidence in adopting such a tool in clinical settings.

In the usability study, a user-centered design approach was employed to establish a reliable interaction between the AI tool and clinicians to unveil their decision-making processes, taking the prediction of postpartum depression (PPD) as a use case to unlock its explainability pathway. This study used a structured environment with pre-defined tasks and assessment criteria for explainability dimensions, as illustrated in the SES metric. Hence, the purpose of this study was valuable in contributing to an in-depth understanding of the explainability of an AI tool and its practical implications for clinicians, ultimately guiding future development.

The research had four phases with a focus on a usability study. The phases included: 1) identification of needs, development and validation of the AI Explainability Framework, 2) development and validation of the SES metric, 3) development, testing, and iterative improvement of the AI prototype, and 4) implementing the usability study involving 20 clinicians.

### A. *AI Explainability Framework development*

Explainability had been widely regarded as a multidimensional construct that enhanced the users' understanding of the model, trust in the model, and ability to effectively utilize an AI tool for clinical decision making. While there were increasing calls for explainable AI-driven systems to demonstrate good governance for public safety in every domain [13], no standardized and widely accepted methods existed to measure the quality of explainability [36]. Although Holzinger et.al. proposed a System Causability Scale [37] to measure the quality of explanations interface or an inherent explanation process, however this scale lacked information related to the development of the questionnaire in terms of validity, reliability, and details about the predictive tool used to evaluate it.

In this study, Explainability was conceptualized as a function of three independent dimensions - Understandability, Trust, and Usability (Fig. 1). Their relationships were expressed as,

Explainability = $f$ (Understandability, Trust, Usability)

This theoretical framework posited that the overarching Explainability arises from the combined and independent contributions of these three dimensions, as illustrated in Fig. 1. Nevertheless, predefined weights were not assigned to any dimension in this framework, allowing the data to empirically reveal the characteristics and strength of these relationships. In this study, the operational definitions of each dimension were provided for better comprehension.

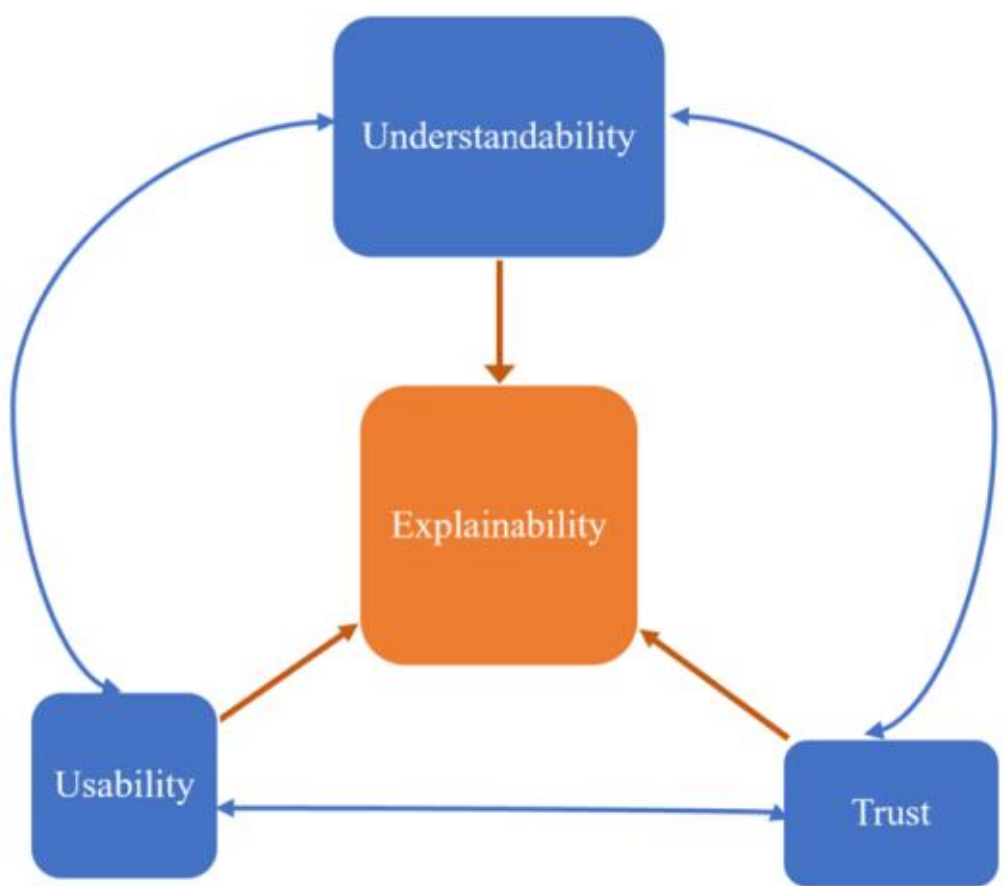

Fig 1. AI Explainability Framework: Connecting Understandability, Trust, and Usability in AI-Driven Tools

**Understandability** refers to the degree to which healthcare providers can comprehend the **input, meaning**, and relevance of the **outputs**, as well as the **decision-making processes,** without the need for extensive technical knowledge.

**Trust** refers to the degree to which healthcare providers have trust in **data security** and **privacy** and **confidence** in the AI methods and their outputs.

**Usability** refers to the degree to which healthcare providers can **interact** with the AI tool seamlessly and its output effectively.

**Explainability** refers to the degree to which healthcare providers can **interact** and **understand** the **input**, and **decision-making** processes, and interpret the **outputs** and **predictions** of the AI tool.

The dimensions of the AI Explainability Framework, as stated above, exhibited notable intersections while possessing unique roles. For instance, the provision of transparent explanations by the AI system could be categorized within either the Understandability or Trust dimensions of the framework. While transparent explanation promoted user understanding, conversely, it bolstered users' confidence in the methodologies. The research team posited that the above system-thinking approach and the proposed operational definitions might help explain how the understandability of the model, trust in the model, and usability of the model-driven tool could influence overarching Explainability.

*B. System Explainability Scale questionnaire development*

*1) Analysis Instrumentation*

Based on the AI Explainability framework, a System Explainability Scale - SES metric (Table 1) was formulated to collect users' perceptions related to Understandability, Trust, and Usability factors through a Likert Scale measurement method. The SES was a 13-item questionnaire (Understandability – 5 questions, Trust – 4 questions, and Usability – 4 questions) with five response options for the users. These options ranged from 'strongly agree' to 'strongly disagree. Each dimension had several questions resulting in a single number that served as a composite measure of the system's overall perceived explainability. The proposed SES metric aimed to rapidly ascertain the overall explainability dimension of the AI tools in the healthcare domain, consequently, determining their suitability for the intended purposes. An additional question was added for users to score the overall explainability of the tool, giving this study the ability to measure the correlation between the composite score of three dimensions with overarching explainability.

This approach gained input from clinicians through a user experimentation (usability study) session, then iterated upon researchers' ideas in a feedback loop mechanism. The session aimed to identify the areas of improvement in the explainability space and refine the tool as necessary. During the early stages of exploration, clinicians were engaged to gather insights into their prevailing challenges in clinical decision making. Such an approach was often used in exploratory research [38], uncovering authentic insights.

TABLE 1. SYSTEM EXPLAINABILITY SCALE (SES) METRIC

| **System Explainability Scale (SES)** | | **Strongly Disagree 1** | **2** | **3** | **4** | **Strongly Agree 5** |
|---|---|---|---|---|---|---|
| **Understandability** (Overall model) | SES1. The tool used clear language, and terminology, avoiding excessive technical jargon | | | | | |
| | SES2. The system's function was straightforward, aiding efficient and error-free task performance | | | | | |
| | SES3. I found the tool organized prediction explanations sequentially for easy understanding | | | | | |
| | SES4. I could easily understand the model's decision process and could interpret its graphics based on the AI model's explainability methods | | | | | |
| | SES5. I could use the resource materials to enhance my awareness and knowledgebase | | | | | |
| **Trust** (Individual prediction results, confidence in method) | SES6. I was confident the tool upheld stringent data privacy and security standards | | | | | |
| | SES7. I had confidence in the data source, methods employed by the tool | | | | | |
| | SES8. I was confident in the reliability of the tool's output in various scenarios | | | | | |
| | SES9. I found the tool transparent in its decision-making, presenting several graphical techniques and interpretation notes regarding method accuracy | | | | | |
| **Usability** | SES10. I found the tool to be user-friendly and efficient, allowing seamless navigation across tabs and features for task completion | | | | | |
| | SES11. I found the tool intuitive in terms of layout and organization of patient input, graphical output, and interpretation notes | | | | | |
| | SES12. I could easily provide feedback and recommendations to refine and improve the tool | | | | | |
| | SES13. I was satisfied with the design and overall performance of the tool | | | | | |
| **System Explainability Scale =** $\sum_{i=1}^{13} \frac{Rating_i}{13 \text{ X } 5}$ | | | | | | |
| **I would give the overall Explainability score of the tool (1-5)** | | | | | | |
| | | *Please enter a number between 1 and 5, inclusive. Fractional numbers are allowed (e.g. 4.8)* | | | | |

$$\text{System Explainability Scale} = \sum_{i=1}^{13} \frac{\text{Rating}_i}{13 \text{ X } 5} \quad (1)$$

Where, $Rating_i$ was the sum of the ratings across all 13 items. $13 \times 5 = 65$, which represented the maximum possible score (if all items received a maximum rating of 5) given by a research participant in Equation (1).

*2) System Explainability Scale questionnaire validation*

The System Explainability Scale (SES) metric questionnaire was evaluated by two independent reviewers (an information scientist and a medical doctor, specialized in maternal fetal medicine) to address the intended objectives of the study. After reviewing the SES questions and their orthogonality according to the conceptual framework, the questionnaire was improved by the lead researcher. In the formative stage, the questionnaire was further validated through employing an internal pilot usability test involving three informaticians with medical doctor (MD) backgrounds.

## *C. AI Prototype development*

A user-centered design methodology was utilized to qualitatively explore clinicians' current needs, challenges, and experiences in managing pregnant people, and to consult medical doctors during the pre-pilot stage in the development of an explainable AI-driven postpartum depression (PPD) risk assessment tool (link: **https://ppd.lairhub.com**/). For readers, user access to the tool was permitted upon request to the lead researcher of this study. A snapshot of the tool was illustrated in Fig. 2. This tool not only classified a patient as having or not having depressive symptoms, but it also presented the probability of PPD with explanations of how the AI tool arrived at its decision. For example, Fig. 2 illustrated how the tool generated a patient-specific risk score (**A**) and provided a list of features that contributed to or mitigated the risk with detailed interpretation notes of the chart (**B**). Additionally, Fig. 2 allowed clinicians to access the top features for all-sample population, enabling them to cross-check the results with individualized risk assessment (**C**). In this example (**C**), the SHAP summary plot was used to display an information-dense summary of how the top features across all-sample population impacted the model's output. By showing the top features across all the population, clinicians could understand which features were generally most influential for PPD predictions. This level of transparency enhanced the clinician's confidence in the model's reliability and accuracy.

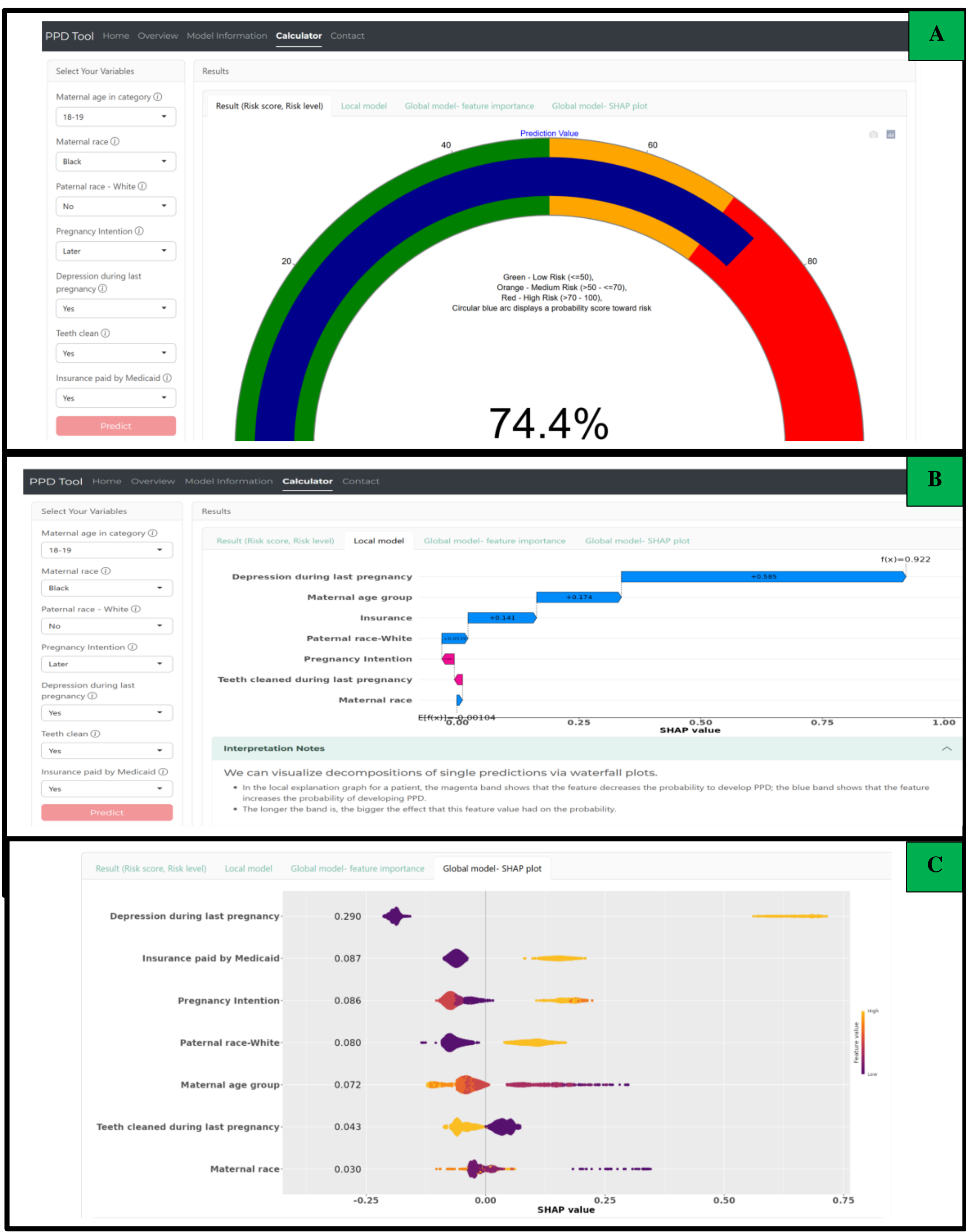

Fig 2. A screenshot of the mock patient record with prediction results (**A**), contributing factors to the risk in the PPD tool (**B**), and contributing factors to the PPD risk in the AI tool for the all-sample population used in the model (**C**).

In summary, the methods used to design the AI prototype (also known as the *PPD tool* in this study) were iterative and implemented over several months. Overall, the prototype development phase included: (1) predevelopment analytic phase to understand user needs and wireframe development, (2) iterative user testing of the prototype itself, and (3) the pilot usability test to perform a review of display, relevance, user control, coherent design and error prevention approach.

## *D. ML Model*

An Extreme Gradient Boosting (XGBoost) classifier model was developed to predict postpartum depression among pregnant individuals using a balanced dataset of 35,518 samples (2016-17) from the Centers for Disease Control and Prevention (CDC) Pregnancy Risk Assessment Monitoring System (PRAMS) survey dataset [39]. The dataset underwent several pre-processing steps, including missing value analysis and the removal of non-informative features with high collinearity or little-to-no variance. The dataset was split into 80% training and 20% testing sets. Linear (logistic regression) and supervised models (k-nearest neighbor, XGBoost and random forest) were tested. XGBoost was finally selected for the performance including speed [40] and ease of cloud deployment as a use case for this study.

XGB helps effectively fit the training data and minimizes the objective functions $L(\theta)$, which consists of a loss function and a penalty term Equation (2) [40], [41]. The loss function $l(.)$ measures how well the predictions $y_{\text{pred}}^{(t)}$ (from the t-th tree) matches with the true labels, $y_{\text{truth.}}$ A regularized penalty is applied to all trees, $f_k$. Each tree, $k$ contributes to the prediction, and its complexity is penalized via the regularization term $\Omega(f_k)$ Equation (3).

$$L(\theta) = \text{Loss term} + \text{Penalty term} \quad (2)$$

$$L(\theta) = \sum_i l\left(y_{\text{pred}}^{(t)}, y_{\text{truth}}\right) + \sum_k \Omega(f_k) \quad (3)$$

Where, $\Omega(f_k) = YT + \frac{1}{2} Y|w|^2$ (4)

In Equation (4),

$Y$ = hyperparameter that controls the penalty for each leaf in the tree, $T$ = number of leaf nodes in the tree,

$w$ = leaf weights

A 10-fold cross validation approach was employed, and optimal hyperparameter were mtry = 1, trees = 300, min_n = 2, tree_depth = 5, learn_rate = 0.0063, loss_reduction = 0.0712, and sample_size = 0.8852. Model performance: ROC-AUC = 0.73, Brier score = 0.24, Sensitivity = 0.58, Specificity = 0.76, Precision = 0.71, Recall = 0.58, and F1-score = 0.64.

XGBoost is a complex model that needs external “post-hoc” techniques to illustrate the explainability in deciding on a specific prediction in the prototype. This type of model cannot assess the significance of each feature on the model’s predictions or their interrelationship. However, there are many local ML explainability methods (e.g., SHapley Additive exPlanations-SHAP) and global model-agnostic methods (e.g., Permutation Feature Importance) that can help explain a Black-box model [22]. In this study, the directional effects of features on the predictions generated by XGBoost were analyzed utilizing SHAP visualizations, which are better aligned with human intuition and can empower users to comprehend the underlying rationale of the model predictions [18]. SHAP represents an interpretability methodology grounded in game theory that aggregates the individual contributions of features in each prediction made by a model Equation (5). Alternative interpretability methods, such as Local Interpretable Model-agnostic Explanations (LIME), offer localized explanations for predictions. However, SHAP was chosen for its capability to comprehensive explanations that encapsulate feature contributions to predictions. SHAP method [18], [22], [42] can present how each feature in the model impacts on the outcome variable (postpartum depression in our study) in the XGBoost model. In simple terms, Shapley values represent the weighted average marginal contribution of a feature to a prediction, taking into account all potential feature coalitions [18], [22] Equation (5).

$$\phi_i = \sum_{S \subseteq F \setminus \{i\}} \frac{|S|!(|F|-|S|-1)!}{|F|!} \left[ f_{S \cup \{i\}}\left(x_{S \cup \{i\}}\right) - f_S(x_S) \right] \quad (5)$$

$\phi_i$ = Shapley value for feature $I$, $F$ = set of all features, $S$ = subset of features that does not include $I$, $|S|$ =number of features in subset $S$, $f_{S \cup \{i\}}$ = prediction of the model when feature $i$ is added to subset $S$, $f(S)$ = prediction of the model using only subset $S$, $|F|!$ = factorial of the total number of features, frac$|S|!(|F|-|S|-1)!|F|!$ = a weight that ensures fairness across all possible combinations of features

The prototype provided the “general information” pertaining to year, data source, and sample size to the clinicians. Furthermore, the details of the developed model were incorporated within the prototype under the tab designated as “model information”, which facilitated transparency and encouraged both acceptance and rejection regarding the efficacy of the model [trust dimension]. The prototype also provided information with literature references to promote the understandability dimension of the clinicians. For instance, detailed interpretation notes were provided on SHAP plot to users as part of an on-demand information service as previously presented in Fig. 2.

The system architecture of the prototype in Fig 2. contained three primary components: (1) user interface using R shiny framework, (2) Machine Learning engine based on tidyverse and tidymodels ecosystems, and (3) web deployment using Git/GitHub, docker image, and Google Cloud platform.

### *E. Usability Study*

The study subjects (n =20) were recruited based on the following criteria. The inclusion criteria included: (1) participants must have clinical background (doctors, nurses or mid-wives) and aged 18 or older, (2) participants must reside in the US and demonstrate fluency in English, both spoken and written, and (3) participants should be able to take part in either in-person or online session for the study. The exclusion criteria were: (1) pregnant women, (2) individuals with potential depressive disorders, and (3) individuals unwilling to be recorded during the session.

Multiple recruitment strategies were employed, including flyer-based, purposive, and snowball sampling methods.

The purposive sampling approach was used to target participants with specialized skills in maternal health, medical informatics, and psychiatry. Additionally, a snowball sampling method was also employed to refer to their eligible colleagues. The above sampling approaches ensured a diverse representation of participants for the usability study. Moreover, no tutorials, user manuals, training on AI tools, visualization techniques, or machine learning methods were provided to participants prior to the study. This approach ensured that all participants interacted with the tool in a consistent and unbiased manner, minimizing the potential familiarity bias. Additionally, participants were not offered compensation for their time. This approach ensured the integrity of the evaluation by reducing potential biases that might arise from compensating participants for their time.

*1) Usability sessions:*

In the summative stage, a comprehensive usability study (a 45-minute remote user testing session) engaging clinicians was conducted using the SES metric which aimed to gather participants' perceptions and satisfaction in interacting with the tool and gauging its overarching explainability characteristics of the AI Tool. This remote moderated usability session via web conference tool (Zoom) was held between July and October 2024, and each session was recorded. Usability study participants used their own computer in a synchronous Zoom session to interact with the tool, sharing screen and controlling the mouse and keyboard.

A "talk-aloud" approach was used to gather feedback while the users reviewed and input patient data and generated predictions with the prototype. The SES metric questionnaire was used to evaluate several aspects of the prototype measured by three dimensions – understandability, trust, and usability.

The participants were also asked to provide the overall perceived explainability score of the prototype. This score enabled the team to assess the correlation of the composite SES score (understandability, trust, and usability) with the overarching explainability score. This approach helped determine if the SES dimensions can be a proxy for the overarching explainability dimension. This study hypothesized that a higher correlation coefficient may demonstrate the utility of the framework for assessing the explainability of an AI tool.

*2) IRB Approval*

In all cases, the lead investigator recorded participants' responses, and the online Qualtrics tool was used for data collection. R statistical software was used for all data analysis and visualizations. The study was approved by the Institutional Review Board (IRB) at the University of North Carolina at Chapel Hill. All participants gave written consent prior to the study.

## III. RESULTS

### A. *Participants Description*

This interpretation assumed that the scale used in the SES metric ranges from a lower value (e.g., 1) to a higher value (e.g., 5), with 5 being the best possible score. Data were analyzed using descriptive statistics, Cronbach's alpha, and the correlation coefficient.

Experimentation sessions were conducted, engaging 20 users (14 female (70%) and six male (30%) participants) to assess the effectiveness, efficiency, and satisfaction of the tool. A set of questionnaires was utilized to facilitate the user testing process. Out of 20 users, 11 (55.00%) had MD degrees, specializing in diverse health domains - maternal and fetal medicine, psychiatry, informatics, etc. Six people (30.00%) had nursing degrees (registered nurse and nurse practitioner), and three (15.00%) had degrees in midwifery. The average age of users was 37.35 years (standard deviation = 7.06 years, max = 53.0 years, min = 28.0 years, median = 38 years, interquartile range = 11 years, skewness = 0.41 and kurtosis = -0.84) and their average professional experience in healthcare after graduation was 10.30 years (standard deviation = 6.17 years, max = 26.0 years, min = 2.0 years, interquartile range = 9.5 years, skewness = 0.62, and kurtosis = -0.16). Only three users (20.00%) had prior experience using an AI/ ML-driven risk prediction tool in their clinical practice. One of these three users had great experience in using such an AI tool in healthcare, whereas the other users thought the tool had a minimum to acceptable levels impact on clinical practice.

Among the respondents, one person (5.00%) reported an advanced level of familiarity with AI-enabled technology, five (25.00%) had an intermediate level, nine (45.00%) had basic familiarity, and five (25.00%) reported no familiarity with the AI-enabled tool. In response to questions on the perceived potential benefits of an AI-driven risk prediction tool in clinical practice, 11 respondents (55.00%) rated significant benefits, six respondents (30.00%) as moderate, and three (15.00%) as very significant.

At the beginning of the experiments, a pre-study questionnaire was provided to gather clinicians' expectations before they accessed the system. The open-ended question "*Which aspects of the risk assessment tool do you think should be prioritized during development?*" was asked to determine their level of interest (Fig. 3). Their responses were synthesized in the context of understandability, trust, and usability dimensions. Most of the participants emphasized the trust dimension of an AI tool and wanted to see that the risk prediction models were primarily reliable and accurate, with a capacity to be generalizable and portable. Participants recommended that the model should use a trustworthy data source that adheres to a privacy-protected protocol. Participants also felt the AI model should be bias-free, including cultural and technological bias. The usability dimension was highlighted by participants, which includes the tool being user-friendly and having a coherent design. The clinicians also wished the tool's output to be easy to understand, allowing the output to be explainable to patients. Participants stated that the system should have an efficient data entry process (minimal number of inputs). One participant highlighted that clinicians should not be over-reliant on AI tools, allowing more focus on in-person care of the patient.

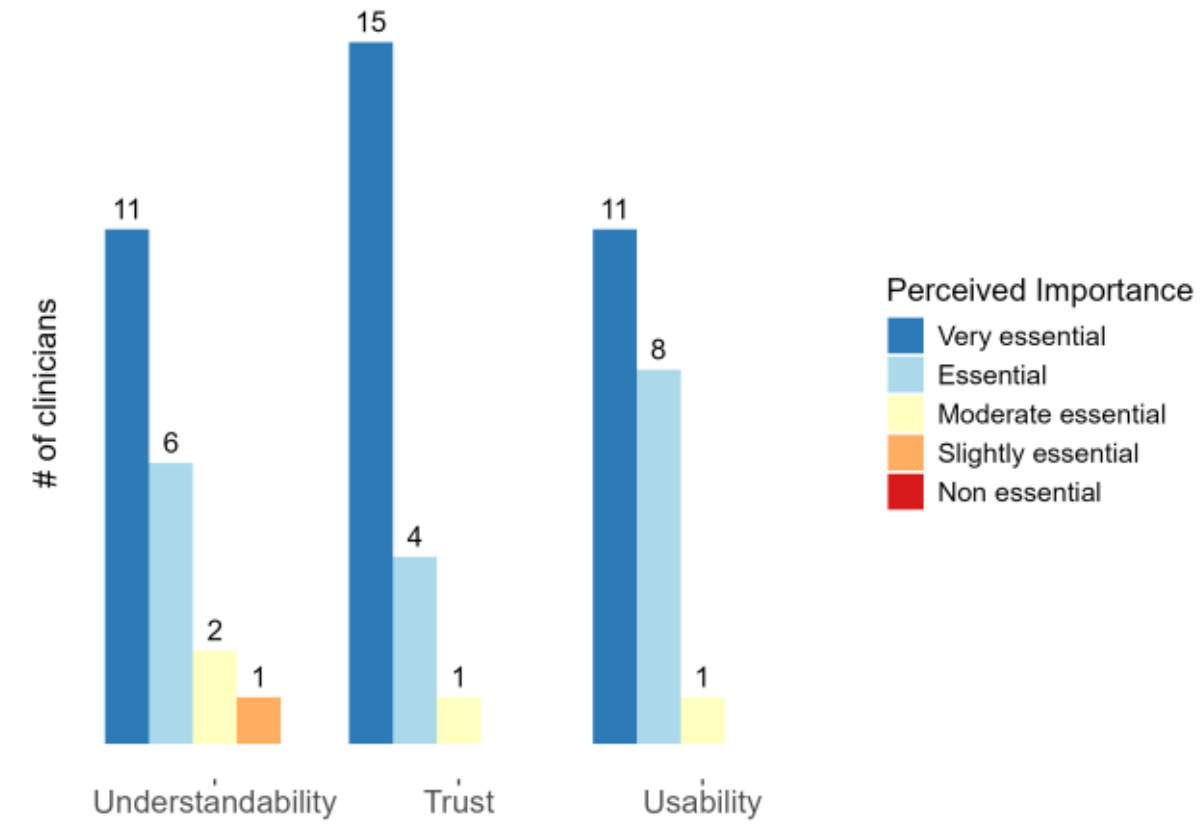

Fig 3. Importance of the explainability dimensions of the SES

### B. *User task performance*

All participants completed 100% of the assigned tasks, which include logging in and reviewing the features, functionalities, and resource materials available in the tool. Individual users created several hypothetical patient scenarios based on their clinical experiences and generated the predictions (probability risk score and level of risk – high, medium, and low) to validate the results with their own assumptions. Participants interrogated the tool to identify the features that contribute to and provide protection from risks. The users did not encounter issues related to errors in producing these results.

### C. *Reliability assessment of the SES metric questionnaire*

The reliability analysis of the 13-item SES score metric indicated strong internal consistency, with a Cronbach's alpha of 0.84 (CI = 0.74 to 0.94 by Duhachek), which was within the acceptance level [43]. The above results indicated that the SES items were robust and satisfactorily measured by the underlying constructs of understandability, trust, and usability, making it a suitable tool for assessing AI explainability.

All participants responded to all the questions of the SES, ensuring a complete dataset for statistical analysis. Across all dimensions of the AI Explainability Framework, the mean score of each dimension was >= 4.50 on a 5-point scale, which indicated that participants were consistently rated the tool as an acceptable AI prototype. The understandability dimension had a mean score of 4.51 (standard deviation = 0.64, skewness = -1.16, kurtosis = 1.17), suggests a concentration of ratings at the higher end with some peak values. The trust dimension achieved a mean score of 4.53 (standard deviation = 0.57, skewness = -0.69, kurtosis = -0.57), demonstrates a relatively uniform dispersion of responses with fewer extreme values. Similarly, the usability dimension exhibited the highest mean score of 4.71 (standard deviation = 0.48, skewness = -1.25, kurtosis = 0.32), presenting a strong clustering of ratings close to the upper limit. The composite average SES score was 4.58 with a standard deviation of 0.58, while the overall explainability score in 4.56 with a low standard deviation (0.46) reflects a strong consensus among participants. These findings demonstrated that participants had a similar level of experiences with the explainability of the prototype.

To further investigate the relationship between the composite SES score and explainability, Spearman's rank correlation coefficient was calculated. The correlation coefficient Spearman's ρ was 0.81, with a p-value of $p<0.001$. This indicated a strong positive relationship between the composite SES score and the explainability score. Higher composite SES scores were associated with higher explainability scores, reflecting a positive nature of the trend (Fig. 4).

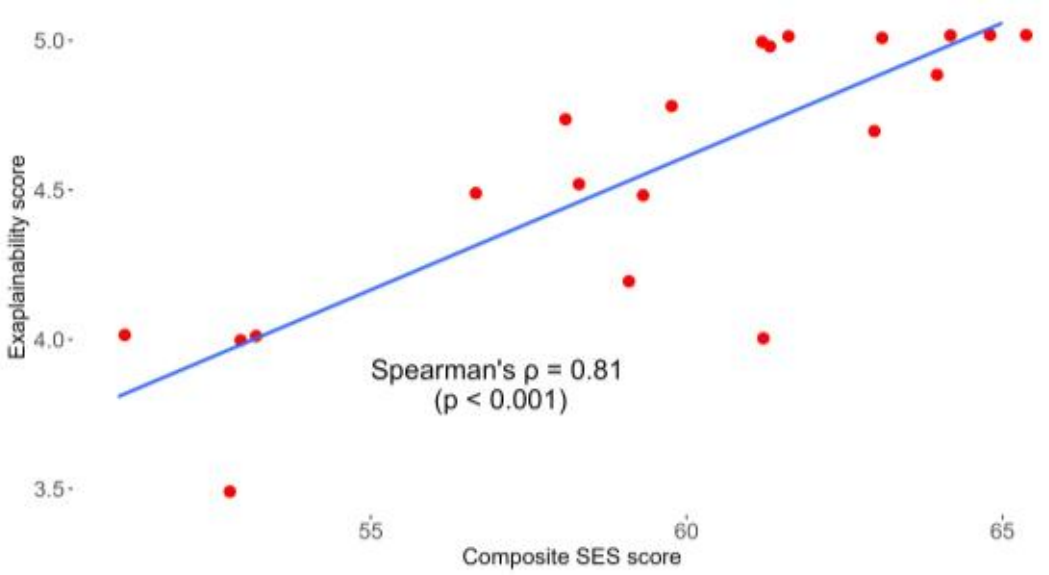

Fig 4. Scatterplot of the correlation between the composite SES score and explainability.

The detailed SES metric showed a consistent level of satisfaction for each element across all participants (Table 2). This analysis attempted to present that understandability, trust, and usability are equivalent to explainability, with the inputs and feedback from 20 clinicians. Past study findings revealed that the Domain Specific-to-context Inspection (DSI) method was most effective for testing users between 12-20 participants, with 91% of usability issues could be identified by only three evaluators [44]. From a usability study perspective, a sample size of 20 participants with relevant experience was acceptable [45], [46].

The participants rated the AI tool's understandability dimension satisfactorily. Participants found the tool's functionalities were efficient and provided error-free task performance (SES2 mean score = 4.80; standard deviation = 0.41). Participants generally agreed that the patient-specific prediction scores and explanations were easy to comprehend (SES3 mean score = 4.60; standard deviation = 0.50). Additionally, the tool was free from excessive technical jargons, which enhanced the participants' understandability of the tool (SES1 mean score = 4.60; standard deviation = 0.60). However, participants had a challenge in quickly grasping the interpretation of the global SHAP summary plot (SES4 mean score = 4.20; standard deviation = 0.70) and emphasized adding more knowledge materials related to the graphics used in this tool (SES5 mean score = 4.35; standard deviation = 0.81).

Participants rated the trust dimension highly on the transparency component of the tool (SES9 mean score = 4.80; standard deviation = 0.41). Participants were also satisfied with privacy and security standards. However, a few participants were not confident enough with the data source (SES7 mean score = 4.35; standard deviation = 0.67), as this study used publicly available survey data to develop the model. The users entered hypothetical patient data and expressed their confidence in the reliability of the tool's prediction score in various scenarios (SES8 mean score = 4.50; standard deviation = 0.51). In terms of usability dimension, the participants found the tool to be intuitive and were satisfied with the design and overall performance of the tool (SES13 mean score = 4.75; standard deviation = 0.44). Participants also appreciated the opportunity to provide feedback to enhance the tool's functions and functionalities (SES12 mean score = 4.80; standard deviation = 0.41) and emphasized the co-design process in developing such a tool. See Table 2 for details.

Table 2. DETAILED SCORES OF SES METRIC

| **System Explainability Scale (SES)** | | **N** | **Responses ( N = 20)** | | | | | **Descriptive Statistics** | | **Shape of Distribution** | |
|---|---|---|---|---|---|---|---|---|---|---|---|
| | | | **Strongly Disagree** | | | | **Strongly Agree** | | | | |
| | | | 1 | 2 | 3 | 4 | 5 | ***Mean*** | ***SD*** | ***Skewness*** | ***Kurtosis*** |
| **Understand ability** (Overall model) | SES1. The tool used clear language, and terminology, avoiding excessive technical jargon | 20 | 0 | 0 | 1 | 6 | 13 | **4.60** | 0.60 | −1.06 | −0.01 |
| | SES2. The system's function was straightforward, aiding efficient and error-free task performance | 20 | 0 | 0 | 0 | 4 | 16 | **4.80** | 0.41 | −1.39 | −0.07 |
| | SES3. I found the tool organized prediction explanations sequentially for easy understanding | 20 | 0 | 0 | 0 | 8 | 12 | **4.60** | 0.50 | −0.38 | −1.95 |
| | SES4. I could easily understand the model's decision process and could interpret its graphics based on the AI model's explainability methods | 20 | 0 | 1 | 0 | 13 | 6 | 4.20 | 0.70 | −1.14 | 2.52 |
| | SES5. I could use the resource materials to enhance my awareness and knowledgebase | 20 | 0 | 0 | 4 | 5 | 11 | 4.35 | 0.81 | −0.66 | −1.24 |
| **Trust** (Individual prediction results, confidence in method) | SES6. I was confident the tool upheld stringent data privacy and security standards | 20 | 0 | 0 | 1 | 9 | 10 | 4.45 | 0.60 | −0.50 | −0.87 |
| | SES7. I had confidence in the data source, methods employed by the tool | 20 | 0 | 0 | 2 | 9 | 9 | 4.35 | 0.67 | −0.47 | −0.93 |
| | SES8. I was confident in the reliability of the tool's output in various scenarios | 20 | 0 | 0 | 0 | 10 | 10 | **4.50** | 0.51 | 0.00 | −2.10 |
| | SES9. I found the tool transparent in its decision-making, presenting several graphical techniques and interpretation notes regarding method accuracy | 20 | 0 | 0 | 0 | 4 | 16 | **4.80** | 0.41 | −1.39 | −0.07 |
| **Usability** | SES10. I found the tool to be user-friendly and efficient, allowing seamless navigation across tabs and features for task completion | 20 | 0 | 0 | 0 | 6 | 14 | **4.70** | 0.47 | −0.81 | −1.41 |
| | SES11. I found the tool intuitive in terms of layout and organization of patient input, graphical output, and interpretation notes | 20 | 0 | 0 | 1 | 6 | 13 | **4.60** | 0.60 | −1.06 | −0.01 |
| | SES12. I could easily provide feedback and recommendations to refine and improve the tool | 20 | 0 | 0 | 0 | 4 | 16 | **4.80** | 0.41 | −1.39 | −0.07 |
| | SES13. I was satisfied with the design and overall performance of the tool | 20 | 0 | 0 | 0 | 5 | 15 | **4.75** | 0.44 | −1.07 | −0.89 |

### D. *Qualitative feedback of the participants*

Clinicians' feedback was gathered based on their verbal comments while working with the AI prototype, both during and after completion of the post-study questionnaire. This data allowed deeper insights into the clinician's cognitive processes and impressions related to the explainability dimensions. The following representative quotes (*as the way feedback was given by the participants*) illustrated the key perspectives of the clinicians on the topic.

One participant noted that "*It's definitely user-friendly. And especially if you have a medical background and you have clinical judgment. It's very easy for me to use, because I understand how all the stuff works. So, like all these, questions are very easy to interpret. Even if you don't have a lot of medical experience. And it's pretty easy to interpret and see how everything correlates, it kind of reminds me, too.*" Another representative statement was: *"It's user friendlier and more accurate. I think it has more criteria, more risk factors. So, I like it.*" A further notable comment was: "*I was surprised by the transparency. I wouldn't have expected that I would not have expected that transparency. And I really like that*." Another participant noted that "*This one will help me see, like, you know, the visual computing like how much it's changing. But this one allows me to see which factor is, is pulling the weight more or less. And so, I think this tool...is also very helpful*"

One participant emphasized the missing part of the tool related to the next steps – "*What can I do now when the risk score hits seventies or eighties - on medications, see a therapist or like, what resources? Yeah, can I connect with now*." In addition, a participant shared: "*More resources could be provided for SHAP plot for the first time user and information and resources on postpartum depression (PPD) could be made available*."

## IV. DISCUSSION AND LIMITATIONS

AI-based clinical decision support tools can improve healthcare quality, but their clinical adoption is still suboptimal due to a lack of clarity about the AI models to the clinicians [47]. From those perspectives, this research is unique in three ways – 1) determining the dimensions of AI explainability in healthcare context and developing an AI Explainability Framework based on three dimensions, 2) developing a measurement method (SES metric) for assessing the explainability within an AI tool, and 3) involving clinicians during the formative and summative phases of the frameworks, prototype development and evaluation. Researchers highlighted the importance of user-centered development over developer-centric approaches as the major success factor for the real-world tool implementation [48]. The explainability measurement of the AI tool is eventually grounded in a usability study that helped this team to understand the perspectives of clinicians and evaluate the tool's features that align with their needs and expectations.

Overall, the tool was rated satisfactorily in the context of understandability dimension for using clear language, providing error-free task execution, and distributing relevant resource materials, with a composite average score of 4.51 (standard deviation = 0.64). The user expressed strong confidence in its reliability and transparency in the decision-making process, with a composite average score of 4.53 (standard deviation = 0.57) under the trust dimension. The usability study found the tool to be user-friendly, intuitive, having adequate provision for a feedback mechanism, and satisfactory in design and performance, with a composite average score of 4.71 (standard deviation = 0.48) reflecting a solid AI prototype. The 13-item SES metric's reliability analysis presented a strong internal consistency and a strong positive relationship between the composite SES score and explainability, indicating a positive trend.

The prototype was considered effective if it accurately predicted the risk of a hypothetical patient, aligning with clinical context. It was deemed efficient for seamless navigation and reduced cognitive load for participants. The information related to model information, data source, tooltip, interpretation notes, and link to additional resource materials reduced the understandability burden and trust issues of the clinicians. Based on the findings, the tool demonstrates that the understandability, trust, and usability of an AI tool are the key dimensions and anchored towards explainability.

Several clinicians highlighted that the tool should trigger patient-specific clinical and behavioral actions based on the probability of risk score and recommended further diagnostics, treatment plans, referrals, or lifestyle modifications. The right to information empowers patients, promotes transparency, and clinicians consider that patients also have the right to know how their health data is utilized and how AI algorithms arrive at specific decisions [49]. For this reason, participants recommended developing a patient-facing app that can facilitate a transparent healthcare journey for patients.

The AI Explainability Framework requires further validation research using the AI prototype, involving more participants as it launches and becomes a more widely used instrument for assessing the AI tools. This study relied on the postpartum depression (PPD) risk prediction tool, which may not be the ideal use case for all clinicians because of their background and clinical practice. Hence, generalizability may suffer to some extent from this study. Although 20 clinicians from medical doctors, nurses, and midwives with diverse areas of clinical practice – psychiatry, infectious disease, and pediatrics were engaged, this study may generalize the need for such a tool and provide feedback for improvements. Nevertheless, the effectiveness of the explainability framework and the relevance of such tools have been established through usability sessions engaging clinicians, possibly leading to future clinical adoption.

## V. CONCLUSIONS AND FUTURE DIRECTIONS

This study demonstrates that understandability, trust, and usability are the three key dimensions of AI Explainability. The framework, measured by the SES metric, provides a rapid and straightforward system-based experimental research methodology to identify and address issues related to explainability barriers for clinicians. In the next iteration of the prototype, this research team will include a feature that triggers patient-specific clinical and behavioral actions based on the probability of risk score, recommending further diagnostics, treatment plans, referrals, or lifestyle modifications. The AI Explainability Framework requires further validation research using the AI prototype, involving more participants to assess the effectiveness of such a tool in clinical settings. Involving clinicians during the planning, design, and experimentation phases may enhance the explainability of the medical AI tools and facilitate a smooth integration into their clinical workflows.

## ACKNOWLEDGEMENT

The researchers would like to thank all mothers who responded to the 2016–2020 PRAMS questionnaire. We thank the PRAMS Working Group, which includes the PRAMS Team, Division of Reproductive Health, CDC, and the PRAMS sites for their role in conducting PRAMS surveillance and allowing the use of their data.

## DISCLAIMER

The findings and conclusions in this report are those of the author(s) and do not necessarily represent the official position of the Centers for Disease Control and Prevention (CDC).

## FUNDING

This project was funded through a grant from the National Center for Advancing Translational Sciences under award number U01TR003629 (PIs: Stuebe, Page, Gurcan; Co-Investigator: Kucirka). The content is solely the responsibility of the authors and does not necessarily represent the official views of the National Center for Advancing Translational Sciences or the National Institutes of Health.